\documentclass[aps,prd,preprint,12pt,superscriptaddress,nofootinbib,a4paper]{revtex4-1}

\usepackage[utf8]{inputenc}
\usepackage{amsmath,amssymb}
\usepackage[separate-uncertainty=true]{siunitx}
\usepackage{graphicx}
\usepackage[usenames,dvipsnames]{xcolor}
\usepackage[caption=false]{subfig}
\usepackage{hyperref}
\usepackage{booktabs}
\usepackage{tabularx}
\usepackage{xspace}
\usepackage{color}
\usepackage[normalem]{ulem}
\usepackage{slashed}
\usepackage{bbold}
\usepackage{wasysym}
\usepackage{graphicx}
\usepackage{mathrsfs} 

\allowdisplaybreaks

\graphicspath{{graphics/}}

\newcommand{\lb}{\left(}
\newcommand{\rb}{\right)}
\newcommand{\be}{\beta}
\newcommand{\al}{\alpha}

\newcommand{\GeV}{{\ensuremath\rm GeV}}
\newcommand{\TeV}{{\ensuremath\rm TeV}}
\newcommand{\lam}{\lambda}

\newcommand{\pb}{{\ensuremath\rm pb}}
\newcommand{\fb}{{\ensuremath\rm fb}}

\DeclareSIUnit{\pb}{pb}
\DeclareSIUnit{\fb}{fb}
\AtBeginDocument{
\heavyrulewidth=.08em
\lightrulewidth=.05em
\cmidrulewidth=.03em
\belowrulesep=.65ex
\belowbottomsep=0pt
\aboverulesep=.4ex
\abovetopsep=0pt
\cmidrulesep=\doublerulesep
\cmidrulekern=.5em
\defaultaddspace=.5em

\newcolumntype{C}{>{\centering\arraybackslash}X}
\newcolumntype{b}{C}
\newcolumntype{s}{>{\hsize=.6\hsize}C}
\newcolumntype{R}{>{\raggedleft\arraybackslash}X}
}

\begin{document}
\date{\today}
\rightline{RBI-ThPhys-2022-14, CERN-TH-2022-068}
\vspace{5mm}
\title{{\Large The THDMa and possible $e^+e^-$ signatures}}

\author{Tania Robens}
\email{trobens@irb.hr}
\affiliation{Ruder Boskovic Institute, Bijenicka cesta 54, 10000 Zagreb, Croatia}
\affiliation{Theoretical Physics Department, CERN, 1211 Geneva 23, Switzerland}

\renewcommand{\abstractname}{\texorpdfstring{\vspace{0.5cm}}{} Abstract}

\begin{abstract}
    \vspace{0.5cm}
 I here discuss the THDMa, a type II two Higgs doublet model that is enhanced by an additional pseudoscalar which serves as a portal to the dark matter sector containing a fermionic dark matter candidate. I present a recent scan of the models parameter space where all parameters are allowed to float freely, and discuss prospects for this model at future $e^+e^-$ colliders for cases that are not covered in standard THDM searches.\\
Talk presented at the 30th International Symposium on Lepton Photon Interactions at High Energies, hosted by the University of Manchester, 10-14 January 2022.
\end{abstract}

\maketitle


\section{Introduction}
I discuss a new physics models that extends the Standard Model (SM) particle sector by an additional scalar and  provides a dark matter candidate. The models is confronted with current theoretical and experimental constraints, including the minimization of the vacuum as well as the requirement of vacuum stability and positivity. I also require perturbative unitarity to hold, and perturbativity of the couplings at the electroweak scale.

From the experimental side, I include the agreement with current measurements of the properties of the 125 \GeV~ resonance discovered by the LHC experiments, as well as agreement with the null-results from searches for additional particles at current or past colliders. Furthermore, I consider bounds from electroweak precision observables (via $S,\,T,\,U$  parameters), 
B-physics observables $\lb B\,\rightarrow\,X_s\,\gamma,\,B_s\,\rightarrow\,\mu^+\,\mu^-,\,\Delta M_s\rb$, as well as agreement with astrophysical observables (relic density and direct detection bounds). In my scan, I use a combination of private and public tools; the latter include HiggsBounds \cite{Bechtle:2020pkv},  HiggsSignals \cite{Bechtle:2020uwn}, SPheno \cite{Porod:2011nf}, Sarah \cite{Staub:2013tta}, micrOMEGAs \cite{Belanger:2018ccd,Belanger:2020gnr}, and MadDM \cite{Ambrogi:2018jqj}. Experimental numbers are taken from \cite{Baak:2014ora,Haller:2018nnx} for electroweak precision observables, \cite{combi} for $B_s\,\rightarrow\,\mu^+\,\mu^-$, \cite{Amhis:2019ckw} for $\Delta M_s$ and \cite{Planck:2018vyg} and  \cite{Aprile:2018dbl} for relic density and direct detection, respectively.
Bounds from $B\,\rightarrow\,X_s\gamma$ are implemented using a fit function from \cite{Misiak:2020vlo,mm}. For predictions of production cross sections I am using  Madgraph5 \cite{Alwall:2011uj}.

\section{THDMa}
The THDMa is a type II two-Higgs-doublet model (THDM), extended by an additional pseudoscalar $a$ mixing with the "standard" pseudoscalar $A$ of the THDM. In the gauge-eigenbasis, the additional field serves as a portal to the dark sector, where I consider a fermionic dark matter candidate $\chi$. This model has extensively been studied in light of hadron colliders, and more details can e.g. be found in \cite{Ipek:2014gua,No:2015xqa,Goncalves:2016iyg,Bauer:2017ota,Tunney:2017yfp,LHCDarkMatterWorkingGroup:2018ufk,Robens:2021lov,Argyropoulos:2022ezr}. 

The model contains, besides the new scalars from the standard THDM, an additional pseudoscalar and the dark matter candidate, leading to the following particle content: ${h,\,H,\,H^\pm}$, ${a,}\,{A,}\,{{\chi}}$. It contains in 12 additional new physics parameters, which can be chosen e.g. as
\begin{eqnarray*}
{v,\,m_h,\,m_H,}\,{ m_a,}\,{m_A,\,m_{H^\pm},}\,{m_\chi};\;{\cos\lb \be-\al\rb,\,\tan\be,}\,{\sin\theta;\;y_\chi,}\,{\lam_3,}\,{\lam_{P_1},\,\lam_{P_2}}.
\end{eqnarray*}
Here $v$ and either $m_h$ or $m_H$ are fixed by current measurements in the electroweak sector. I refer the reader to \cite{Robens:2021lov} for a more thorough discussion, including the concrete form of the potential.

\begin{center}
\begin{figure}
\begin{center}
\begin{minipage}{0.45\textwidth}
\begin{center}
\includegraphics[width=\textwidth]{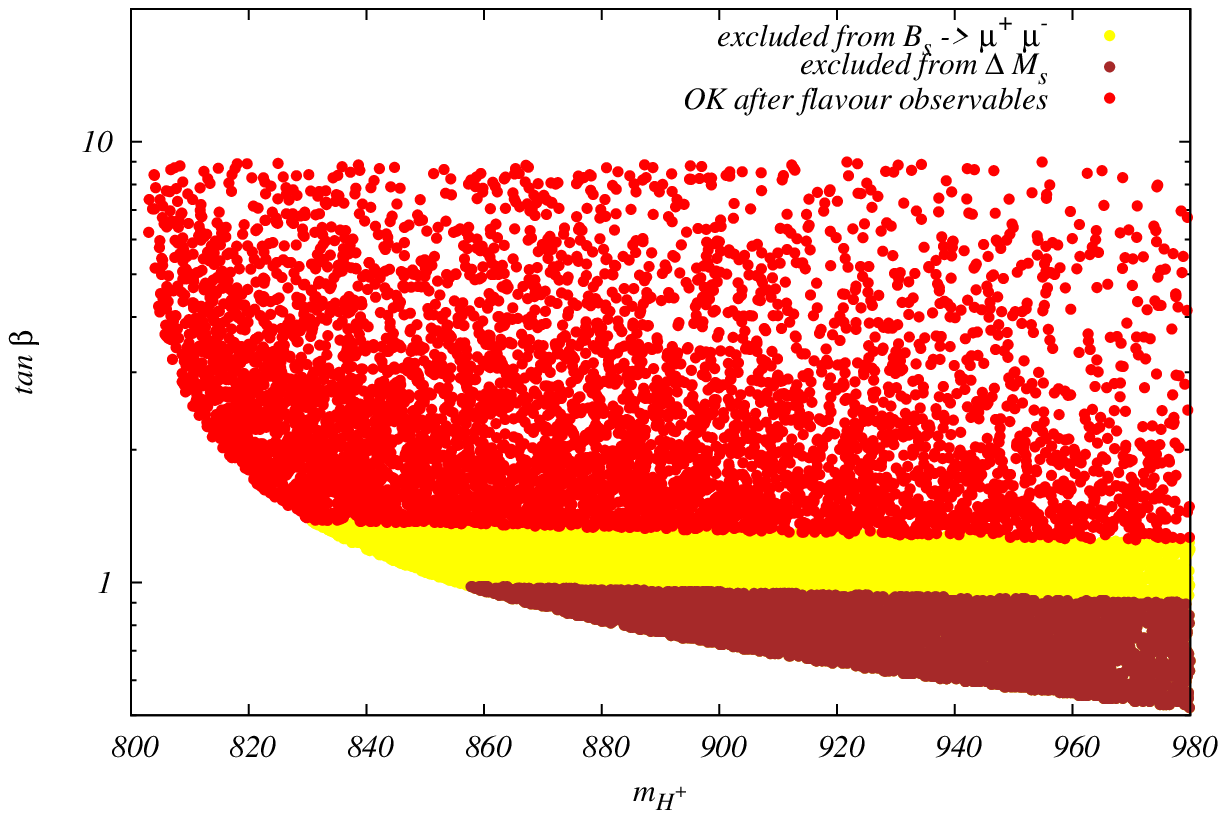}
\end{center}
\end{minipage}
\begin{minipage}{0.45\textwidth}
\begin{center}
\includegraphics[width=\textwidth]{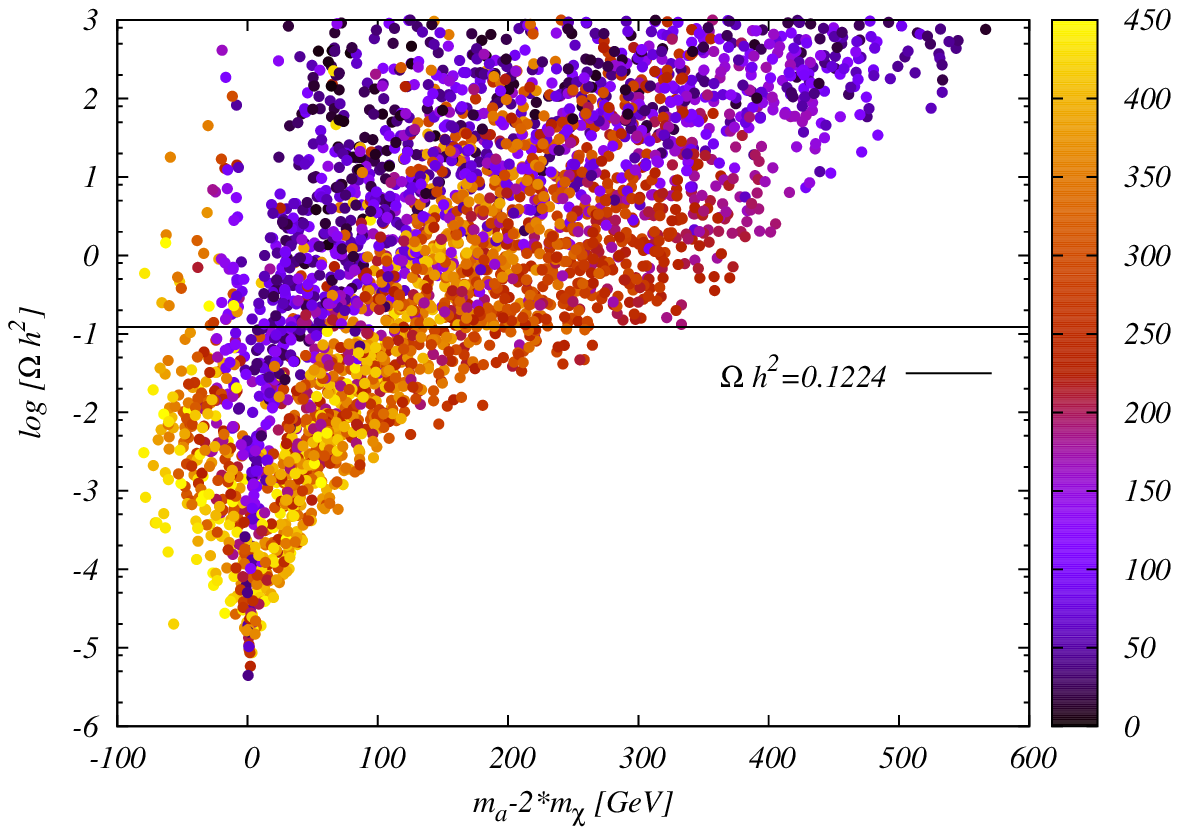}
\end{center}
\end{minipage}
\end{center}
\caption{\label{fig:thdmab} {\sl Left:} Bounds on the $\lb m_{H^\pm},\,\tan\be\rb$ plane from B-physics observables. The contour for low $\lb m_{H^\pm,\,\tan\be}\rb$ values stems from \cite{Misiak:2020vlo,mm}. {\sl Right:} Dark matter relic density as a function of $m_a-2\,m_\chi$, where $m_\chi$ defines the color coding. The typical resonance-enhanced relic density annihilation is clearly visible. Figures taken from \cite{Robens:2021lov}.}
\end{figure}
\end{center}

I here report on results of a scan that allows all of the above novel parameters float in specific predefined ranges \cite{Robens:2021lov}. Due to the large number of free parameters, it is not always straightforward to display bounds from specific constraints in 2-dimensional planes. In some cases, however, displaying these in such setups is straightforward. Two examples are shown in figure \ref{fig:thdmab}. The first plot shows bounds in the $\lb m_{H^\pm},\,\tan\be \rb$ plane from B-physics observables. The result is similar to a simple THDM, and shows that in general low masses $m_{H^\pm}\lesssim\,800\,\GeV$  as well as values $\tan\be\lesssim\,1$ are excluded. The second plot displays the relic density as a function of the mass difference $m_a-2\,m_\chi$. Here, a behaviour can be observed that is typical in many models with dark matter candidates: in the region where this mass difference remains small, relic density annihilates sufficiently to stay below the observed relic density bound, leading to a so-called "funnel" region. Too large differences lead to values $\Omega\,h_c\,\gtrsim\,0.12$ and therefore are forbidden from dark matter considerations.

\begin{center}
\begin{figure}
\includegraphics[width=0.67\textwidth]{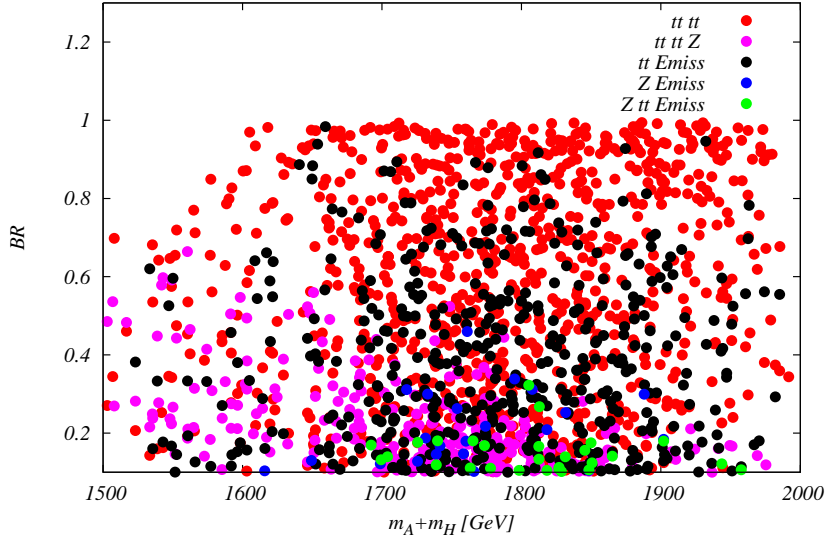}
\caption{\label{fig:brsAH} Branching ratios into various final states for $A\,H$ production, as a function of the mass sum.}
\end{figure}
\end{center}

\begin{center}
\begin{figure}
\begin{center}
\begin{minipage}{0.45\textwidth}
\begin{center}
\includegraphics[width=\textwidth]{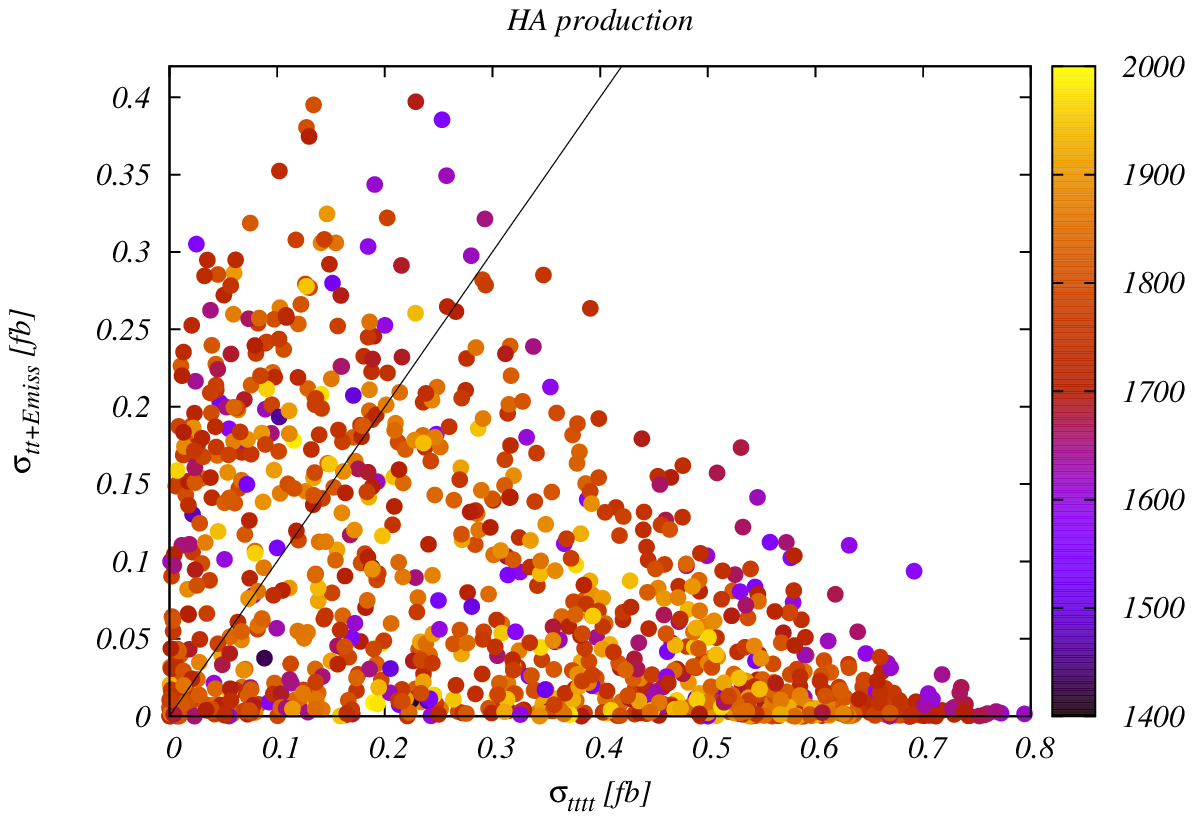}
\end{center}
\end{minipage}
\begin{minipage}{0.45\textwidth}
\begin{center}
\includegraphics[width=\textwidth]{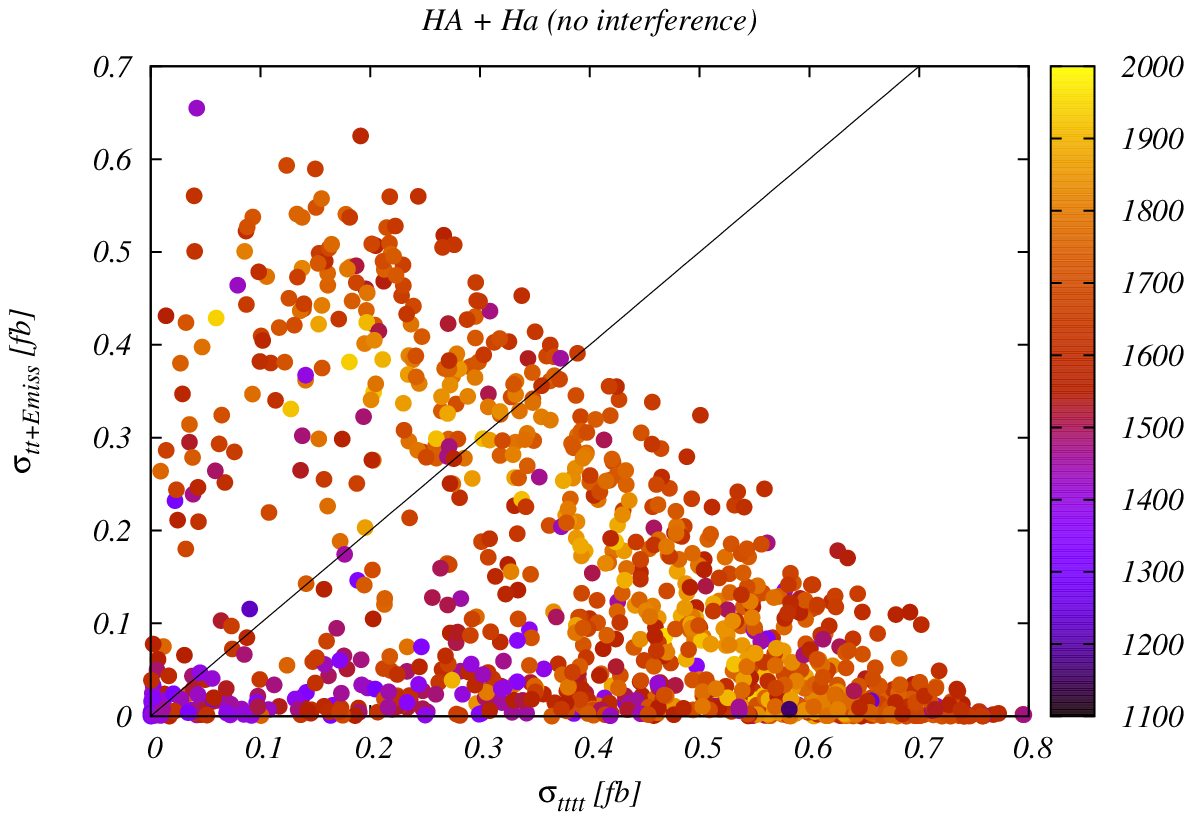}
\end{center}
\end{minipage}
\end{center}
\caption{\label{fig:thdmaatee} Production cross sections for $t\bar{t}t\bar{t}$ (x-axis) and $t\bar{t}+\slashed{E}$ (y-axis) final state in a factorized approach at 3 \TeV~ center-of-mass energy. {\sl Left:} mediated via $HA$, {\sl right:} mediated via $HA$ and $Ha$ intermediate states. Color coding refers to $m_H+m_A$ {\sl (left)} and $m_H+0.5\times\,\lb m_A+m_a\rb$ {\sl(right)}. Figures taken from \cite{Robens:2021lov}.}
\end{figure}
\end{center}

Finally, it is interesting to consider which production cross-section values would still be feasible for points that fulfill all constraints \cite{Robens:2021lov} at $e^+e^-$ colliders. I concentrate on signatures that include missing energy and therefore do not exist in a THDM without a portal to the dark sector. Processes that include the lighter CP-even scalar, as e.g. $e^+e^-\,\rightarrow\,hA, ha$ are supressed due to alignment, which makes $e^+e^-\,\rightarrow\,HA, Ha$ the most interesting channel that contains novel signatures. Due to the interplay of B-physics and electroweak constraints, such points typically have mass scales $\gtrsim\,1\,\TeV$. Therefore, the first interesting scenario are production cross sections for an $e^+e^-$ collider with a center-of-mass energy of 3 \TeV. The corresponding production cross sections are shown in figure \ref{fig:thdmaatee}. Here, I display predictions for $t\,\bar{t}\,t\,\bar{t}$ and $t\,\bar{t}+\slashed{E}$ final states using a factorized approach. There is a non-negligible number of points where the second channel is dominant. A "best" point with a large rate for $t\,\bar{t}+\slashed{E}_\perp$ has been presented in \cite{Robens:2021lov}:

\begin{eqnarray}
&&\sin\theta\,=\,-0.626,\;\cos\lb \be-\al\rb\,=\,0.0027,\,\tan\be\,=\,3.55 \nonumber\\
&&m_H\,=\,643\,\GeV,\,m_A\,=\,907\,\GeV,\,m_{H^\pm}\,=\,814\,\GeV, \nonumber\\
&&m_a\,\,=\,653\,\GeV,\,m_\chi\,=\,277\,\GeV, \nonumber\\
&&y_\chi\,\,=\,-1.73,\,\lambda_{P_1}\,=\,0.18,\,\lambda_{P_2}\,=\,2.98,\,\lambda_3\,=\,8.63.
\end{eqnarray}
For this point, all width/ mass ratios are $\lesssim\,6\,\%$. In addition, branching ratios for various final states as a function of the mass sum for the $HA$ channel are given in figure \ref{fig:brsAH}.

\section{Conclusion and Outlook}

In this work, I presented a model that extend the particle content of the SM and also provides a dark matter candidate. In particular, I defined signatures that do not exist in models without dark matter candidates. I have presented production cross sections for various standard pair-production modes. A more dedicated investigation of the corresponding signatures, including background simulation and cut optimization, is in the line of future work.
\section*{Acknowledgements}
This research was supported in parts by the National Science Centre, Poland, the HARMONIA
project under contract UMO-2015/18/M/ST2/00518 (2016-2021), OPUS project under contract
UMO-2017/25/B/ST2/00496 (2018-2021).
%

\end{document}